\documentclass[varenna]{cimento}

\usepackage[utf8]{inputenc}
\usepackage{graphics,epsfig}
\usepackage{amsmath}
\usepackage{collref}

\def\beq{\begin{equation}}   \def\eeq{\end{equation}}
\def\bea{\begin{eqnarray}}   \def\eea{\end{eqnarray}}

\title{Enhanced Higgs-Mediated Lepton-Flavour-Violating\\Processes in the Supersymmetric Inverse Seesaw Model}
\shorttitle{Higgs Mediated LFV in SUSY Inverse Seesaw}

\author{C. Weiland*}

\institute{Laboratoire de Physique Théorique d'Orsay, CNRS--UMR 8627, Université Paris Sud 11\\Bâtiment 210, F--91405 Orsay Cedex, France\\ *E-mail: Cedric.Weiland@th.u-psud.fr}

\shortauthor{C. Weiland}

\begin{document}

\maketitle

\begin{abstract}
We study the impact of the inverse seesaw mechanism on several leptonic and hadronic low-energy flavour-violating observables in the context of the Minimal Supersymmetric Standard Model. Indeed, the contributions of the light right-handed sneutrinos from the inverse seesaw significantly enhance the Higgs-mediated penguin diagrams. We find that this can increase the different branching ratios by as much as two orders of magnitude.
\end{abstract}

\newpage

\section{Introduction}

Being highly suppressed in the Standard Model (SM), any charged lepton-flavour-violating (cLFV) signal would be a clear evidence of new physics: mixings in the lepton sector and probably the presence of new particles, possibly shedding light on the origin of neutrino mass generation. Among the extensions of the SM, supersymmetry (SUSY)  is a well-motivated solution for the hierarchy problem,  providing also gauge couplings unification and dark matter candidates. However, the minimal supersymmetric extension of the SM (MSSM) does not provide a mechanism to generate neutrino masses, which can be done by embedding a seesaw mechanism in the MSSM. The inverse seesaw \cite{Mohapatra:1986bd} is a very appealing choice since it provides large neutrino Yukawa couplings  ($Y_\nu \sim O(1)$) and a seesaw scale close to the electroweak one, thus within LHC reach.

\vspace*{-2mm}
\section{Supersymmetric Inverse Seesaw and Higgs-Mediated LFV}

The SUSY inverse seesaw model consists of the MSSM extended by three pairs of singlet superfields, $\widehat{\nu}^c_i$ and $\widehat{X}_i$ ($i=1,2,3$),
with opposite lepton numbers, and is defined by the superpotential
\vspace*{-4mm}
\bea
{\mathcal W}&=& \varepsilon_{ab} \left[
Y^{ij}_d \widehat{D}_i \widehat{Q}_j^b  \widehat{H}_d^a
              +Y^{ij}_{u}  \widehat{U}_i \widehat{Q}_j^a \widehat{H}_u^b 
              + Y^{ij}_e \widehat{E}_i \widehat{L}_j^b  \widehat{H}_d^a + Y^{ij}_\nu 
\widehat{\nu}^c_i \widehat{L}^a_j \widehat{H}_u^b - \mu \widehat{H}_d^a \widehat{H}_u^b \right] \nonumber \\
              &+& M_{R_i}\widehat{\nu}^c_i\widehat{X}_i+
\frac{1}{2}\mu_{X_i}\widehat{X}_i\widehat{X}_i  ~,
\label{eq:SuperPot}
\eea
\vspace*{-6mm}

\noindent where $i,j = 1,2,3$ denote generation indices.
In the above,  $\widehat H_d$ and $\widehat H_u$ are the down- and
up-type Higgs superfields, $\widehat L_i$ denotes the SU(2)
doublet lepton superfields.
$M_{R_i}$ represents the right-handed (RH) neutrino
mass term which conserves lepton number while $\mu_{X_i}$ violates it. More details are given in \cite{Abada:2011hm}. In this model, the smallness of the left-handed neutrino mass is due to the smallness of $\mu_{X_i}$, thus leaving the mass parameters $m_D= Y_\nu v_u$ and $M_R$ unconstrained. The only source of flavour violation is encoded in 
the neutrino Yukawa couplings $Y_\nu$. Even under the assumption of 
universal soft breaking terms at the GUT scale, radiative effects will induce flavour violation in the slepton mass 
matrices, which in turn gives rise to slepton-mediated cLFV \cite{Borzumati:1986qx, Hisano:1995cp, Hisano:1998fj}. 
As an example, the RGE 
corrections to the left-handed slepton soft-breaking masses are given by
\vspace*{-4mm}
\begin{equation}
(\Delta m_{\widetilde{L}}^2)_{ij}\simeq
-\frac{1}{8\pi^2}(3m_0^2+A_0^2) 
(Y_\nu^\dagger L Y_\nu)_{ij}=\xi (Y^\dagger_\nu Y_\nu)_{ij} \,, ~~ L=\ln\frac{M_{GUT}}{M_{R}} \,.
\label{slepmixing}
\end{equation}
\vspace*{-6mm}

\begin{figure}
\vspace*{-4mm}
\begin{minipage}{0.52\textwidth}
\begin{center}
\epsfig{file=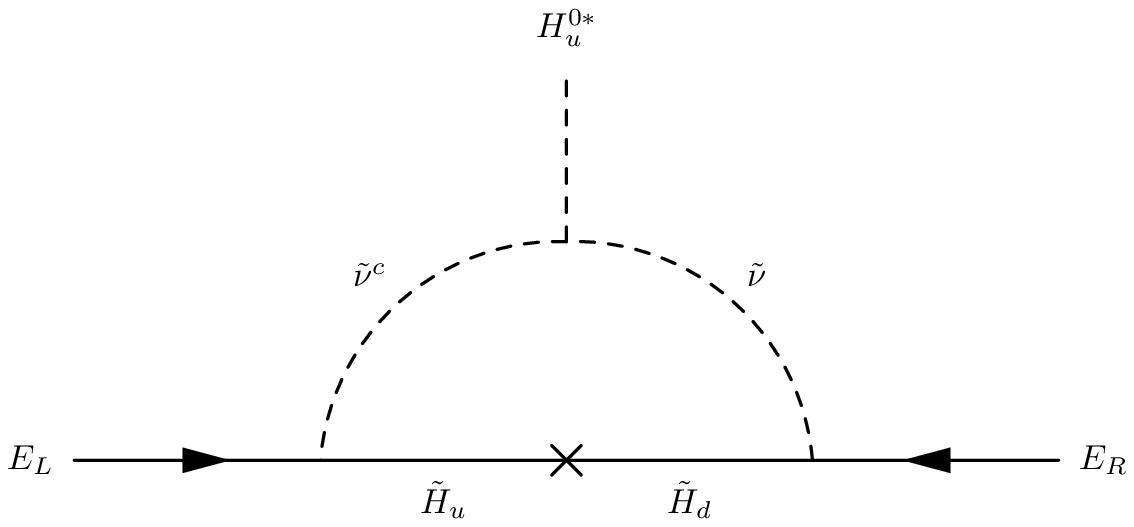, width=70mm}
\caption{Right-handed sneutrino contribution  to $\epsilon'_2$.}\label{2}
\end{center}
\end{minipage}
\begin{minipage}{0.47\textwidth}
 \begin{center}
  \epsfig{file=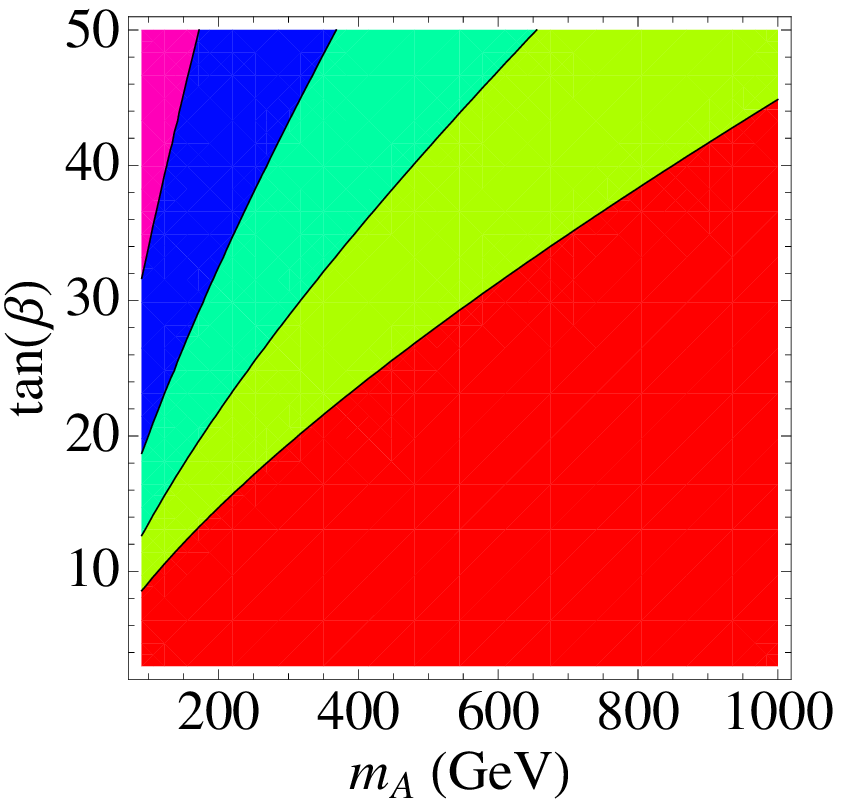, width=60mm}
\caption{$\text{Br}(\tau \rightarrow 3 \mu)$, the contours corresponding, from left to right, to $2.1 \times 10^{-8}$, $10^{-9}$, $10^{-10}$, $10^{-11}$.}\label{3}
 \end{center}
\end{minipage}
\vspace*{-8mm}
\end{figure}

Compared to the type I seesaw, where $M_{R}\sim10^{14}$ GeV, the inverse seesaw has a RH neutrino mass scale $M_{R}\sim\mathcal{O}(\text{TeV})$ and this leads to an enhancement of the factor $\xi$  and hence to all low-energy cLFV observables. Furthermore, having RH sneutrinos with a mass $M^2_{\widetilde \nu^c} \sim M^2_\text{SUSY}$, 
the $\widetilde \nu^c$-mediated processes are no longer suppressed, and 
might even significantly contribute to the low-energy LFV observables through the new diagram given in fig. \ref{2}. Notice that in the type I SUSY seesaw, this contribution is usually neglected.

The effective Lagrangian describing the couplings of the neutral Higgs fields
to the charged leptons being given by \cite{Babu:2002et}
\vspace*{-3mm}
\bea
-{\cal L}^\text{eff}=\bar E^i_R Y_{e}^{ii} \left[ 
\delta_{ij} H_d^0 + \left(\epsilon_1 \delta_{ij} + 
\epsilon_{2ij} (Y_\nu^\dagger Y_\nu)_{ij} \right) H_u^{0\ast }
\right] E^j_L + \text{h.c.}  \,, 
\label{Leff}
\eea
\vspace*{-7mm}

\noindent the new contribution due to the diagram of fig. \ref{2} can be expressed as
\vspace*{-3mm}
\bea 
\epsilon'_{2ij}= - \mu A_\nu \frac{\mu^2 m^2_{\widetilde \nu_i}\ln (\mu^2/m^2_{\widetilde \nu_i})+m^2_{\widetilde \nu_i} m^2_{\widetilde \nu^c_j}\ln (m^2_{\widetilde \nu_i}/m^2_{\widetilde \nu^c_j})+m^2_{\widetilde \nu^c_j} \mu^2\ln (m^2_{\widetilde \nu^c_j}/\mu^2)}
{16\pi^2(\mu^2-m^2_{\widetilde \nu_i})(m^2_{\widetilde \nu_i}-m^2_{\widetilde \nu^c_j})(m^2_{\widetilde \nu^c_j}-\mu^2)} \,.\label{newdiag}
\eea
\vspace*{-5mm}

\noindent Explicit formulas for the various coefficients in \ref{Leff} are given and evaluated in \cite{Abada:2011hm}. The main feature is that, in the SUSY inverse seesaw, this new diagram provides the dominant Higgs-mediated contribution and, together with the increase of the factor $\xi$, it enhances $\epsilon_2$ by a factor of order $\sim 10$ compared to the standard seesaw.

\vspace*{-2mm}
\section{Results and discussion}

We studied several cLFV observables dominated by the Higgs penguin diagrams and found that they would be enhanced by as much as two orders of magnitude compared to the type I SUSY seesaw. It is interesting to note that $m_A$ and $\tan \beta$ are the most relevant parameters
in the Higgs-mediated flavour violating processes. To better illustrate this, in fig. \ref{3} we study the dependence of Br($\tau \rightarrow 3 \mu$) on the aforementioned parameters. We have chosen three points (CMSSM-A and CMSSM-B correspond to the 10.2.2 and 40.1.1 benchmark points in \cite{AbdusSalam:2011fc}, NUHM-C is an example of a non-universal scenario) that are summarized in table \ref{tab:sfp10.1} and we calculated several branching ratios in table \ref{4}.

\begin{table}[htb]
\vspace*{-4mm}
\begin{center}
\begin{tabular}{lrrrrrrrr}
\hline
Point & $\tan\beta$ & $m_{1/2}$ & $m_0$ & $m^2_{H_U}$& $m^2_{H_D}$ &$A_0$  &$\mu$&$m_A$ \\
\hline
CMSSM-A &10& 550 & 225 & $(225)^2$ &$(225)^2$ &0 &690 & 782 \\
CMSSM-B &40& 500 & 330 & $(330)^2$&$(330)^2$ &-500&698 &604 \\
NUHM-C &15& 550 & 225 &$(652)^2$ &$-(570)^2$ &0&478&150  \\
\hline
\end{tabular}
\vspace*{-3mm}
\caption{Benchmark points used in the numerical analysis (dimensionful parameters in GeV).\label{tab:sfp10.1} }
\vspace*{-1mm}
 \begin{tabular}{crrlll}
  \hline
    LFV Process & Present Bound & Future Sensitivity & CMSSM-A & CMSSM-B & NUHM-C\\
  \hline
    $\tau \rightarrow \mu \mu \mu$ & $2.1\times10^{-8}$\cite{Hayasaka2010139} & $8.2 \times 10^{-10}$\cite{OLeary2010af} & $1.4 \times 10^{-15}$ & $3.9 \times 10^{-11}$ & $8.0 \times 10^{-12}$ \\
    $\mu \rightarrow e e e$ &  $1.0 \times 10^{-12}$\cite{Bellgardt:1987du} &  & $6.3 \times 10^{-22}$ & $1.5 \times 10^{-17}$ & $3.7 \times 10^{-18}$ \\
    $\tau \rightarrow \mu \eta$ & $2.3\times 10^{-8}$\cite{collaboration:2010ipa} & $\sim 10^{-10}$\cite{OLeary2010af} & $8.0 \times 10^{-15}$ & $3.3 \times 10^{-10}$ & $4.6 \times 10^{-11}$ \\
    $B^{0}_{d} \rightarrow \mu \tau$ & $2.2\times 10^{-5}$\cite{Aubert:2008cu} & & $2.7 \times 10^{-15}$ & $8.5 \times 10^{-10}$ & $2.7 \times 10^{-11}$ \\
    $B^{0}_{s} \rightarrow e \mu$ & $2.0\times 10^{-7}$\cite{Aaltonen:2009vr} & $6.5\times 10^{-8}$\cite{Bonivento:1028132} & $3.4 \times 10^{-16}$ & $8.9 \times 10^{-11}$ & $3.4 \times 10^{-12}$ \\
    $h \rightarrow \mu \tau$ & & & $1.3 \times 10^{-8}$ & $2.6 \times 10^{-7}$ & $2.3 \times 10^{-6}$\\
  \hline
 \end{tabular}
\end{center}
\vspace*{-5mm}
  \caption{Higgs-mediated contributions to the branching ratios of several LFV
processes with current experimental bounds and future sensitivities.}\label{4}
\vspace*{-8mm}
\end{table}

One can see that from an experimental point of view, 
the most promising channels in the SUSY inverse seesaw are 
$\tau \rightarrow \mu \mu \mu$ and $\tau \rightarrow \mu \eta$ 
which could be tested at the next generation of $B$ factories.
It is important to stress that the numerical results summarised in  table \ref{4} correspond to considering 
{\it only} Higgs-mediated contributions. For large $\tan\beta$ values, Higgs penguins provide the leading contributions. However, at small $\tan \beta$, due to the sizeable contributions from photon and Z penguins, our results should be interpreted as partial contributions. Another interesting property of the Higgs-mediated processes is that the corresponding amplitude strongly depends on the chirality of the heaviest lepton, which can induce an asymmetry that potentially allows to identify if 
 Higgs mediation is the dominant contribution to the LFV observables.

\vspace*{-2mm}
\section{Conclusion}
In this work, we have studied Higgs-mediated LFV processes  in the framework of the SUSY inverse seesaw. We found that TeV scale RH sneutrinos enhance the Higgs-mediated contributions of several LFV processes by as much as two orders of magnitude when compared to the standard (type I) SUSY seesaw. 

\vspace*{-1mm}
\subsection*{Acknowledgements}
The author acknowledges financial support from the Societ\`a Italiana de Fisica to attend the School. This work has been partly done under the ANR project CPV-LFV-LHC NT09-508531.

\vspace*{-4mm}
\bibliographystyle{varenna}
\bibliography{References}

\end{document}